\documentclass{PoS}

\makeatletter

\def\compoundrel#1\over#2{\mathpalette\compoundreL{{#1}\over{#2}}}
\def\compoundreL#1#2{\compoundREL#1#2}
\def\compoundREL#1#2\over#3{\mathrel
	{\vcenter{\hbox{$\m@th\buildrel{#1#2}\over{#1#3}$}}}}
\makeatother
\newcommand{\bfi}[1]{\mbox{\boldmath $#1$}}
\usepackage{graphicx}

\title{
Baryon-baryon interaction of strangeness $\bfi{\bf S=-1}$ sector
}

\ShortTitle{
Baryon-baryon interaction of strangeness $S=-1$ sector
}

\author{\speaker{Hidekatsu Nemura}
        \thanks{Present address: Center of Computational Sciences,
        University of Tsukuba, Tsukuba, Ibaraki, 305-8577, Japan}
	\\
	Department of Physics,
	Tohoku University,
	Sendai,
	980-8578, Japan \\

        E-mail: \email{nemura.hidekatsu.gb@u.tsukuba.ac.jp}}

\author{
        for HAL QCD %
	Collaboration\\
	\includegraphics[width=0.20\textwidth]{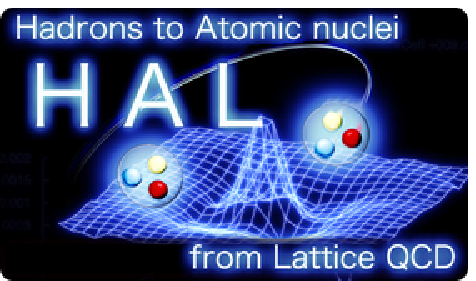}
	}

\abstract{
We present our recent studies on hyperon-nucleon ($YN$) 
interactions in the strangeness $S=-1$ that $p\Lambda, \Sigma^0 p$ and $\Sigma^+ n$, 
by extracting corresponding potentials through  Nambu-Bethe-Salpeter wave functions. 
We calculate $\Lambda N$ and $\Sigma N$ potentials in the isospin $I=3/2$ channel, 
using the $N_f=2+1$ gauge configurations generated by PACS-CS collaboration
and employing an improved method to obtain potentials in lattice QCD simulations. 
For the $^1S_0$ channel, 
the central $\Sigma N (I=3/2, ^1S_0)$ potential and the central $\Lambda N (^1S_0)$ 
potential are found to be very similar.
In the spin triplet ($^3S_1-^3D_1$) channels, 
the central $\Lambda N(^3S_1-^3D_1)$ potential is attractive 
while the central $\Sigma N(I=3/2, ^3S_1-^3D_1)$ potentials is 
repulsive.
Tensor potentials, on the other hand, are rather weak in the diagonal
part of both $\Lambda N$ and  $\Sigma N(I=3/2)$
systems.
}

\FullConference{ The XXIX International Symposium on Lattice Field Theory - Lattice 2011\\
July 10-16, 2011\\
Squaw Valley, Lake Tahoe, California}

\begin{document}

\section{Introduction}

The $\Lambda$-nucleon ($\Lambda N$) and 
the $\Sigma$-nucleon ($\Sigma N$) interactions are one of 
the basic inputs to study the hypernuclear systems. 
in which hyperons (or strange quarks) are embedded 
in normal nuclear systems as ``impurities''~\cite{Hashimoto:2006aw}. 
For example,  from  studies of few-body systems for $s$-shell $\Lambda$-hypernuclei,
it has been pointed out that 
a coupled-channel $\Lambda N-\Sigma N$ interaction plays 
a significant role to make a hypernucleus being bounded.\cite{Nemura:2002fu}. 
On the other hand, 
no $\Sigma$-hypernuclei have been observed except for 
four-body $\Sigma$-hypernucleus ($^4_\Sigma$He), and furthermore 
a recent experimental study suggests that 
$\Sigma$-nucleus interaction is repulsive. 
Such informations are useful to study  properties of hyperonic matters inside the 
neutron stars~\cite{SchaffnerBielich:2008kb},
though a hyperonic equations of state (EOS) employed in such a study may contradict
 a recent observation of a  massive neutron star heavier than $2M_{\odot}$~\cite{Demorest2010}. 
Despite their importance,
present phenomenological 
 $\Lambda N$ and $\Sigma N$ interactions have still 
large uncertainties since direct $\Lambda N$ and $\Sigma N$ 
   scattering experiments are either difficult or 
   impossible due to the short life-time of hyperons. 

Recently, a new  approach has been proposed lattice QCD to study 
not only the $NN$ interaction but also various baryonic 
interactions including the H-dibaryon 
system~\cite{Ishii:2006ec,Aoki:2009ji,Murano:2011nz,Nemura0806.1094,Inoue:2010hs,Inoue:2010es,Doi:2011gq}, and
the method is recently extended to systems in inelastic channels~\cite{Aoki:2011gt,kenjisLAT2011}. 
This method has been applied also to study various interactions 
other than baryon-baryon interactions~\cite{Ikeda:2011qm,Ikeda:2011bs}.  
In this approach, 
the interparticle potential can be directly extracted in lattice QCD 
through the Nambu-Bethe-Salpeter (NBS) wave function, and observables 
 such as the phase shift and the binding energy  can be calculated 
  by using the resultant potentials. 
The purpose of this report is to present 
our recent calculations of the 
$\Lambda N$ potentials 
as well as the $\Sigma N (I=3/2)$ potentials 
using full QCD gauge configurations. 
Several earlier results had already been reported 
at LATTICE 2008\cite{Nemura:2009kc} and LATTICE 2009\cite{Nemura:2010nh}. 
This report is the latest version of those reports, which includes 
several new results: 
(i) $\Sigma N (I=3/2)$ potentials are studied. 
(ii) An improved method is employed to extract potentials more precisely~\cite{Ishii:2011LAT}.

\section{Improved extraction of potentials}

In the HAL QCD scheme, the non-local but energy-independent potential is defined  
by the  Schr\"{o}dinger equation as
\begin{equation}
 (\vec{\nabla}^2 + k^2)~\phi(\vec{r}) =
  2\mu 
  \int  d^3r^\prime U(\vec{r},\vec{r}^\prime)
  \phi(\vec{r}^\prime),
\end{equation}
where $\phi(\vec{r})$ is the equal-time NBS wave function of two-baryon system ($B_1,B_2$),
 $\mu=m_{B_1}m_{B_2}/(m_{B_1}+m_{B_2})$ and 
$k^2$ %
are the reduced mass of the $(B_1, B_2)$ %
system and the square of asymptotic momentum %
in the center-of-mass frame, respectively. 
In practice, the nonlocal potential is expanded in terms of the velocity(derivative) as~\cite{TW67},
$
U(\vec{r},\vec{r}^\prime)=
 V_{B_1 B_2}(\vec{r},\vec{\nabla})\delta(\vec{r}-\vec{r}^\prime).
$
The potential $V$ may have an
antisymmetric spin-orbit term 
when two baryons are not identical. 
For example, the $\Lambda N$ potential is given by 
\begin{eqnarray}
 V_{\Lambda N} &=&
  V_0(r)
  +V_\sigma(r)(\vec{\sigma}_{\Lambda}\cdot\vec{\sigma}_{N})
  +V_T(r)S_{12}
  +V_{LS}(r)(\vec{L}\cdot\vec{S}_+)
  +V_{ALS}(r)(\vec{L}\cdot\vec{S}_-)
  +{O}(\nabla^2),
  \label{GenePotNL}
\end{eqnarray}
where
$S_{12}=3(\vec{\sigma}_{\Lambda}\cdot\vec{n})(\vec{\sigma}_{N}\cdot\vec{n})-\vec{\sigma}_{\Lambda}\cdot\vec{\sigma}_{N}$ 
is the tensor operator with $\vec{n}=\vec{r}/|\vec{r}|$, 
$\vec{S}_{\pm}=(\vec{\sigma}_{N} \pm \vec{\sigma}_{\Lambda})/2$  are %
symmetric ($+$) and antisymmetric ($-$) spin operators, 
$\vec{L}=-i\vec{r}\times\vec{\nabla}$ is the orbital %
angular momentum operator. %
In the velocity expansion,
$V_{0,\sigma,T}$ are the leading order (LO) potentials, denoted by $V_{\rm LO}$,
while $V_{LS,ALS}$ are of the next-to-leading-order (NLO). 
In this report, we consider  the LO potentials only. 

In lattice QCD simulations
we first calculate the normalized four-point correlation function defined by
\begin{equation}
 { R}_{\alpha\beta}^{(J,M)}(\vec{r},t-t_0) = 
 \sum_{\vec{X}}
 \left\langle 0
  \left|
   B_{1,\alpha}(\vec{X}+\vec{r},t)
   B_{2,\beta}(\vec{X},t)
   \overline{{\cal J}_{B_{3} B_{4}}^{(J,M)}(t_0)}
  \right| 0 
 \right\rangle
 /
 \exp\{
 -(m_{B_1}+m_{B_2})(t-t_0)
 \},
\end{equation}
where the summation over $\vec{X}$ selects  states with 
zero total momentum. 
The $B_{1,\alpha}(x)$ and $B_{2,\beta}(y)$ denote the 
interpolating fields of the baryons such as 
\begin{eqnarray}
 &&
  p = \varepsilon_{abc} \left(
			 u_a C\gamma_5 d_b
			\right) u_c,
  \qquad \quad ~
  n = - \varepsilon_{abc} \left(
			   u_a C\gamma_5 d_b
			  \right) d_c,
  \\
 &&
  \Sigma^{+} = - \varepsilon_{abc} \left(
				    u_a C\gamma_5 s_b
				   \right) u_c,
  \qquad
  \Sigma^{0} = {1\over\sqrt{2}} \varepsilon_{abc} \left\{
						   \left(
						    d_a C\gamma_5 s_b
						   \right) u_c
						   +
						   \left(
						    u_a C\gamma_5 s_b
						   \right) d_c
						  \right\},
  \\
 &&
  \Lambda = {1\over \sqrt{6}} \varepsilon_{abc}
  \left\{
   \left(
    d_a C\gamma_5 s_b
   \right) u_c
   +
   \left(
    s_a C\gamma_5 u_b
   \right) d_c
   - 2
   \left(
    u_a C\gamma_5 d_b
   \right) s_c
  \right\},
\end{eqnarray}
and
$\overline{{\cal J}_{B_3B_4}^{(J,M)}(t_0)}=
  \sum_{\alpha^\prime\beta^\prime}
  P_{\alpha^\prime\beta^\prime}^{(J,M)}
  \overline{B_{3,\alpha^\prime}(t_0)}
  \overline{B_{4,\beta^\prime}(t_0)}$
is a source operator which creates $B_3B_4$ states with the
total angular momentum $J,M$. 
This normalized four-point function can be expressed as
\begin{equation}
 \begin{array}{lll}
  { R}_{\alpha\beta}^{(J,M)}(\vec{r},t-t_0) &=&
   \sum_{n} A_{n}
   \sum_{\vec{X}}
   \left\langle 0
    \left|
     B_{1,\alpha}(\vec{X}+\vec{r},t)
     B_{2,\beta}(\vec{X},t)
    \right| E_{n} 
   \right\rangle
   {\rm e}^{-(E_{n}-m_{B_1}-m_{B_2})(t-t_0)},
 \end{array}
\end{equation} 
where $E_n$ ($|E_n\rangle$) is the eigen-energy (eigen-state)
of the six-quark system 
with the particular quantum number
(i.e., %
$J^\pi,M$,
strangeness $S$ and
isospin $I$), 
and 
$A_n = \sum_{\alpha^\prime\beta^\prime} P_{\alpha^\prime\beta^\prime}^{(JM)}
\langle E_n | \overline{B}_{4,\beta^\prime}
\overline{B}_{3,\alpha^\prime} | 0 \rangle$. 

Since $E_n - m_{B_1}-m_{B_2} = k^2/(2\mu) +O(k^4)$,
we have\cite{Ishii:2011LAT}
\begin{eqnarray}
\left(\frac{\nabla^2}{2\mu} -\frac{\partial}{\partial t}\right){ R}(\vec r,t)
&=& \int d^3r^\prime\, U(\vec r,\vec r^\prime){ R}(\vec r^\prime,t) +O(k^4)
=V_{\rm LO}(\vec r) { R}(\vec r,t) +\cdots ,
\end{eqnarray}
where 
$t-t_0$ should be moderately large so that states with large $k^2$ and states with more than 2 particles are suppressed. 
%
%

For the spin
singlet
state, we extract the %
central potential as 
$V_C(r;J=0)=({\nabla^2\over 2\mu}-{\partial\over \partial t})
{ R}/{ R}$. 
For the spin triplet state, %
the wave function 
is decomposed into 
the $S$- 
and 
the $D$-wave components as
\begin{equation}
 \left\{
 \begin{array}{l}
  R_{\alpha\beta}(\vec{r};\ ^3S_1)={\cal P}R_{\alpha\beta}(\vec{r};J=1)
   \equiv {1\over 24} \sum_{{\cal R}\in{ O}} {\cal R}
   R_{\alpha\beta}(\vec{r};J=1),
   \\
  R_{\alpha\beta}(\vec{r};\ ^3D_1)={\cal Q}R_{\alpha\beta}(\vec{r};J=1)
   \equiv (1-{\cal P})R_{\alpha\beta}(\vec{r};J=1).
 \end{array}
 \right.%
\end{equation}
Therefore, 
the Schr\"{o}dinger equation with the LO 
potentials for the spin triplet state becomes
\begin{equation}
 \left\{
 \begin{array}{c}
  {\cal P} \\
  {\cal Q}
 \end{array}
 \right\}
 \times
 \left\{
  -{\nabla^2\over 2\mu} 
  +V_0(r)
  +V_\sigma(r)(\vec{\sigma}_{\Lambda}\cdot\vec{\sigma}_{N})
  +V_T(r)S_{12}
 \right\}
 { R}(\vec{r},t-t_0)
 =-
 \left\{
 \begin{array}{c}
  {\cal P} \\
  {\cal Q}
 \end{array}
 \right\}
 \times
 {\partial \over \partial t}~
 { R}(\vec{r},t-t_0),
\end{equation}
from which
the 
central and the tensor potentials, 
$V_C(r;J=0)=V_0(r)-3V_\sigma(r)$ for $J=0$, 
$V_C(r;J=1)=V_0(r) +V_\sigma(r)$, and $V_T(r)$ for $J=1$, can be
determined. 
%


\section{Numerical simulations}

In our calculations for $\Lambda N$ and $\Sigma N (I=3/2)$ potentials,
we employ
2+1 flavor full QCD gauge configurations generated by 
PACS-CS collaboration~\cite{Aoki:2008sm} 
with the RG-improved Iwasaki gauge action and 
the nonperturbatively $O(a)$-improved 
Wilson quark action at $\beta=1.9$ on 
a $L^3\times T=32^3\times 64$ lattice.  
The lattice spacing at the physical quark masses is
$a=0.0907(13)$ fm~\cite{Aoki:2008sm}, 
thus the spatial volume is $(2.90 \mbox{fm})^3$. 
We have chosen one set of hopping parameters 
$(\kappa_{ud},\kappa_s)=(0.13700, 0.13640)$ 
for light and the strange quarks, which correspond to 
$(m_\pi,m_K)\cong (699,787)$ MeV. 
The wall source  with 
the Coulomb gauge fixing is placed at the time-slice $t_0$,
and the Dirichlet boundary condition is imposed 
in the temporal direction at the time-slice $t-t_0=N_t/2$. 
A total number of  gauge configurations we have used is  399,
and we put a source at $t_0=8n$ with $n=0,1,2,\cdots, 8$ 
on each configuration to increase the statistics. 

\section{Results}
%


\subsection{$\Lambda N$ and $\Sigma N(I=3/2)$ potentials
in $^1S_0$ channel} 
\begin{figure}[t]
 \centering \leavevmode
 \begin{minipage}[t]{0.49\textwidth}
  \includegraphics[width=.99\textwidth]
  {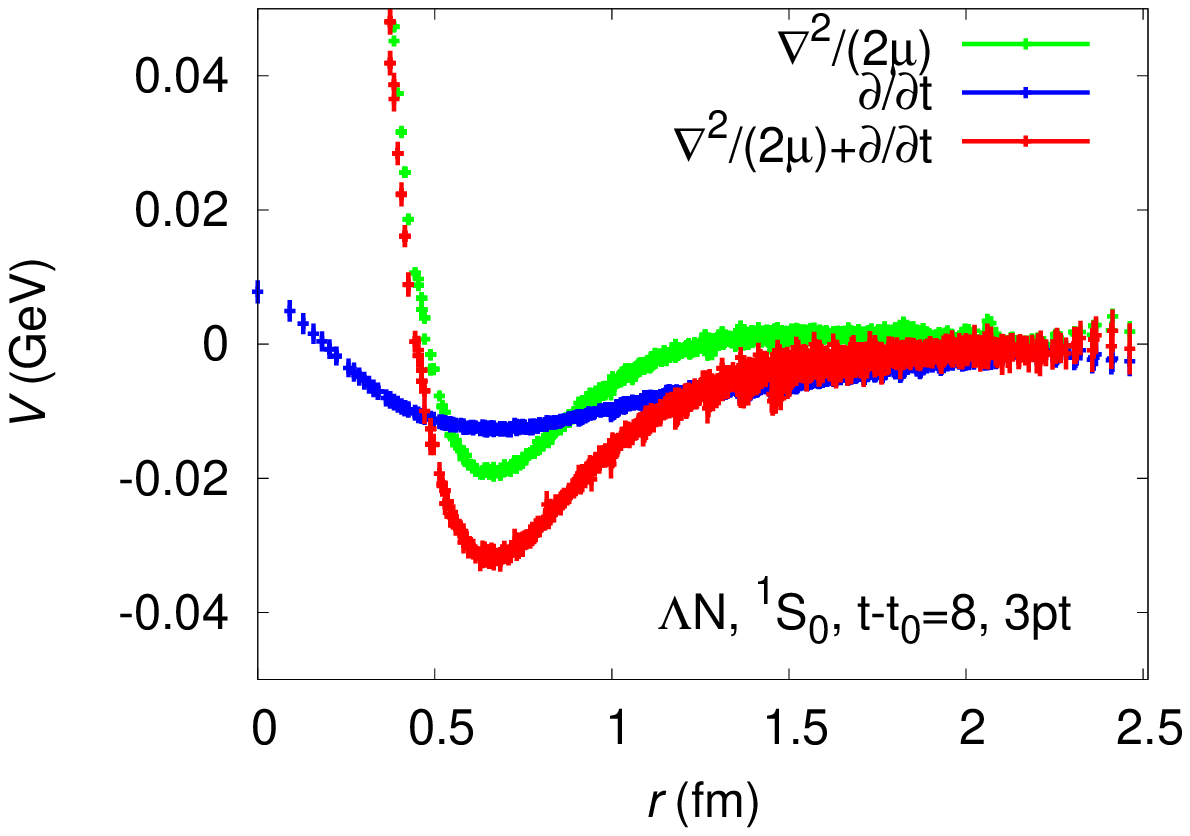}
  \footnotesize
 \end{minipage}~
 \begin{minipage}[t]{0.49\textwidth}
  \includegraphics[width=.99\textwidth]
  {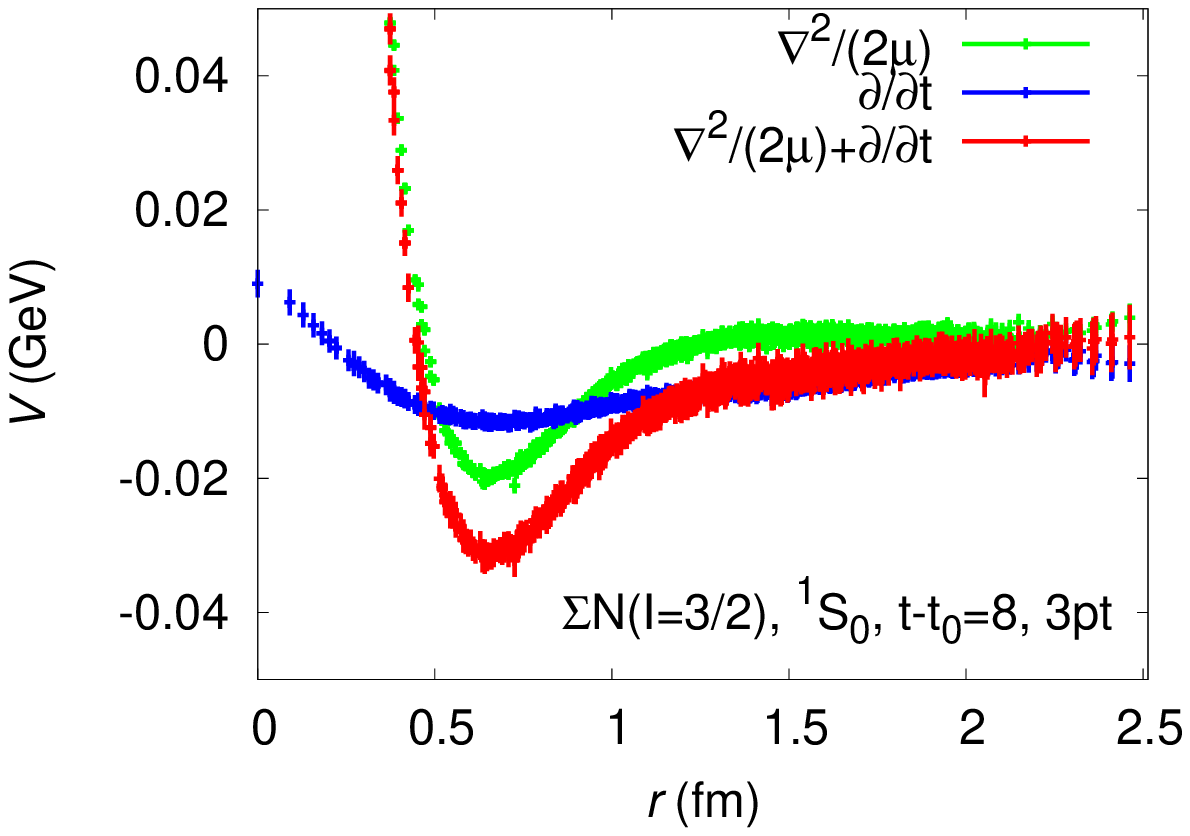}
 \end{minipage}
 \caption{
 Left:
 The central potential (red) 
 in the $^1S_0$ channel of the $\Lambda N$ system
 in $2+1$ flavor QCD as a function of $r$, together with 
 the ${\mathbf{\nabla}^2\over 2\mu}$ part (green),
 ${\partial\over \partial t}$ part (blue).
 Right:
  The central potential 
 in the $^1S_0$ channel of the $\Sigma N(I=3/2)$ system
 as a function of $r$.
 Symbols are same as the left figure.
 }
 \label{LN_and_SN2I3_VC_1S0}
\end{figure}
The $\Lambda N$ (left panel) and the 
$\Sigma N(I=3/2)$ (right panel) potentials 
in the $^1S_0$ channel are shown in
Figure~\ref{LN_and_SN2I3_VC_1S0}, where
the laplacian part is represented by green, the time-derivative part by blue and
the total potential by red, respectively.
Note that the time derivative term gives a significant attraction
at medium and longer distances. 

In the 2+1 flavor QCD, 
while the $\Sigma N$ ($I=3/2$) potential still belongs directly 
to the $\mathbf{27}(I=3/2)$ representation thanks to the isospin($I$) symmetry,
an energy eigenstate of a $\Lambda N$ system in the $^1S_0$ channel 
 is a mixture of $\mathbf{27} (I=1/2)$ and $\mathbf{8}_s$ in the flavor representation,
so that these two potentials are not necessarily mutually in agreement. 
As seen from Figure~\ref{LN_and_SN2I3_VC_1S0}, however,
these two potentials look very similar to each other,
since the flavor symmetry breaking is still small 
in the present $2+1$ flavor QCD calculation, where 
the baryon masses are given by
$(m_N,m_\Lambda,m_\Sigma)=\left(1.574(3),1.635(3),1.650(3)\right)$~GeV. 

\subsection{$\Lambda N$ interaction in the $^3S_1-^3D_1$ channel}
\begin{figure}[t]
 \centering \leavevmode
 \begin{minipage}[t]{0.49\textwidth}
  \includegraphics[width=.99\textwidth]
  {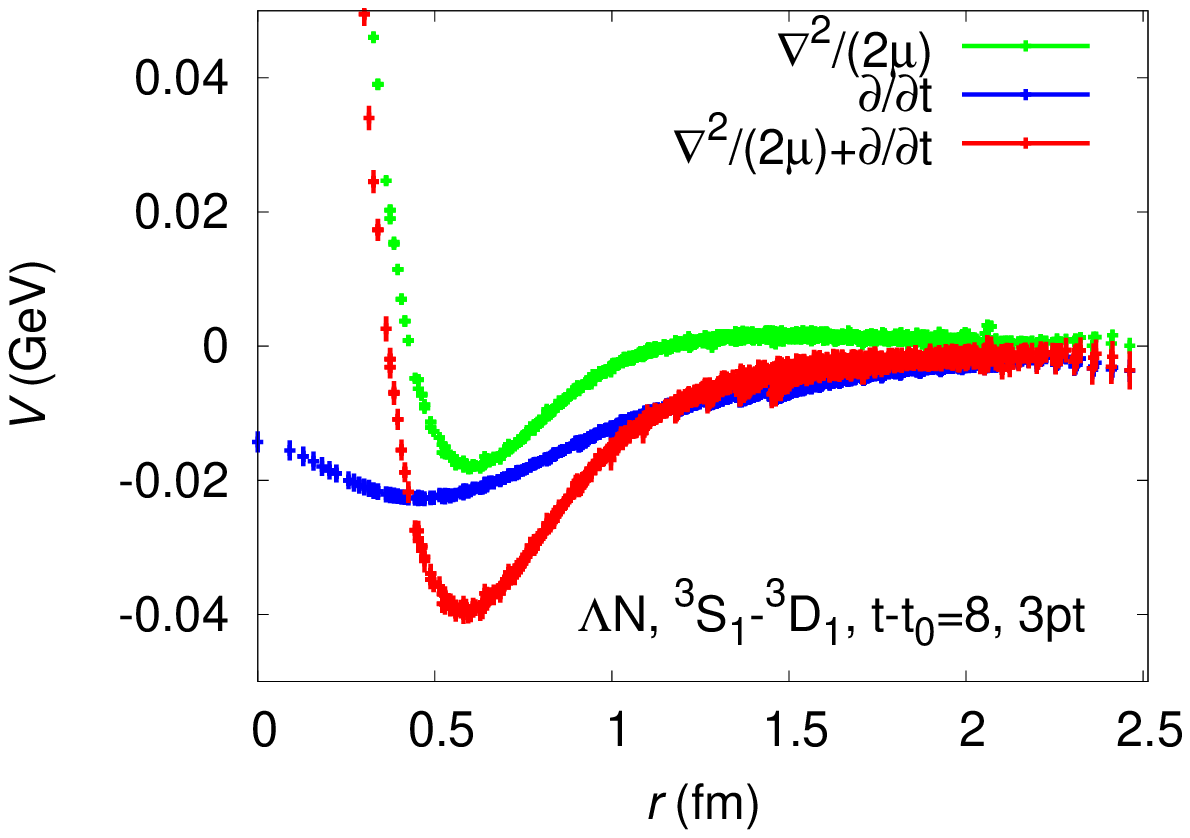}
  \footnotesize
 \end{minipage}
 \hfill
 \begin{minipage}[t]{0.49\textwidth}
  \includegraphics[width=.99\textwidth]
  {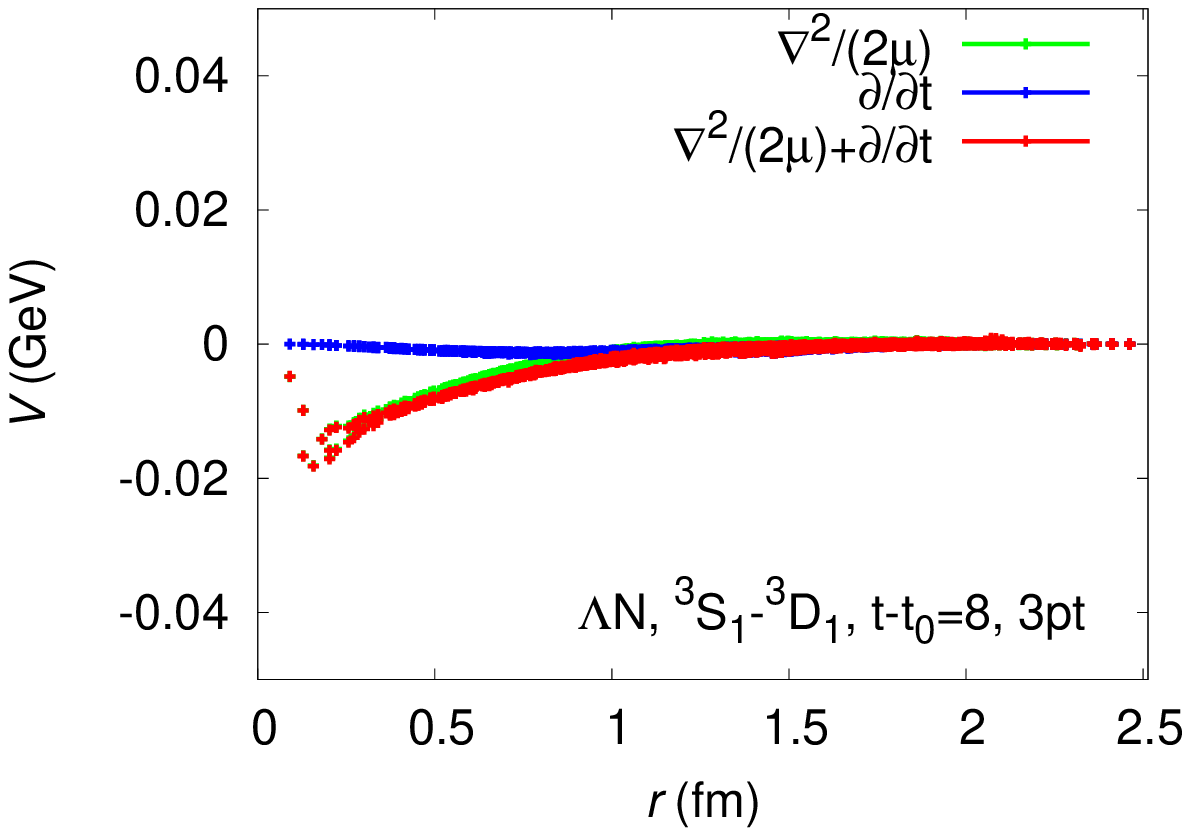}
 \end{minipage}
 \caption{
 Left:
 The central potential in the $^3S_1-^3D_1$ channel of the 
 $\Lambda N$ system as a function of $r$. 
Symbols are same as the Fig.~\protect\ref{LN_and_SN2I3_VC_1S0}. 
 Right:
  The tensor potential in the $^3S_1-^3D_1$ channel of the
 $\Lambda N$ system as a function of $r$. 
 Symbols are
 same as Fig.~\protect\ref{LN_and_SN2I3_VC_1S0}.
}
 \label{LN_VCT_3E1}
\end{figure}
Figure~\ref{LN_VCT_3E1} shows the central potential (left panel) 
and the tensor potential (right panel) of the $\Lambda N$ system 
in the $^3S_1-^3D_1$ channel, 
whose eigenstate is a mixture of $\overline{\mathbf{10}}$ and $\mathbf{8}_a$.
The time-derivative part (blue) of the central potential has attractive 
contributions, so that the attraction of the total potential (red) is enhanced
at long distance. 
The attractive well at the distance $r\approx 0.6$~fm 
is deeper than that of the $\Lambda N$ central potential in the $^1S_0$ channel.
The time derivative part of the tensor potential, on the other hand, 
gives a negligible contribution. Furthermore, 
the tensor potential itself (red) is weaker than the tensor potential 
of the $NN$ system\cite{Ishii:2011LAT}. 

\subsection{$\Sigma N$ interaction in the $^3S_1-^3D_1$ channel}
\begin{figure}[t]
 \centering \leavevmode
 \begin{minipage}[t]{0.49\textwidth}
  \includegraphics[width=.99\textwidth]
  {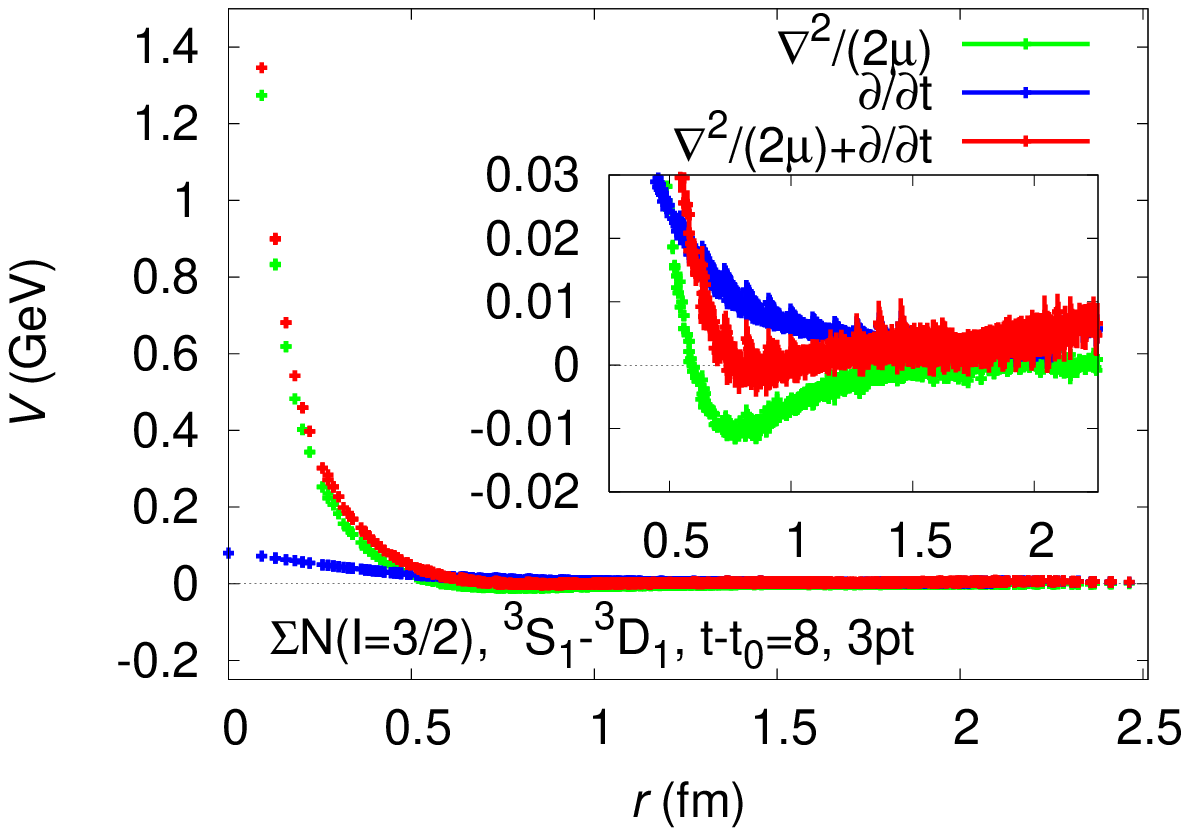}
  \footnotesize
 \end{minipage}
 \hfill
 \begin{minipage}[t]{0.49\textwidth}
  \includegraphics[width=.99\textwidth]
  {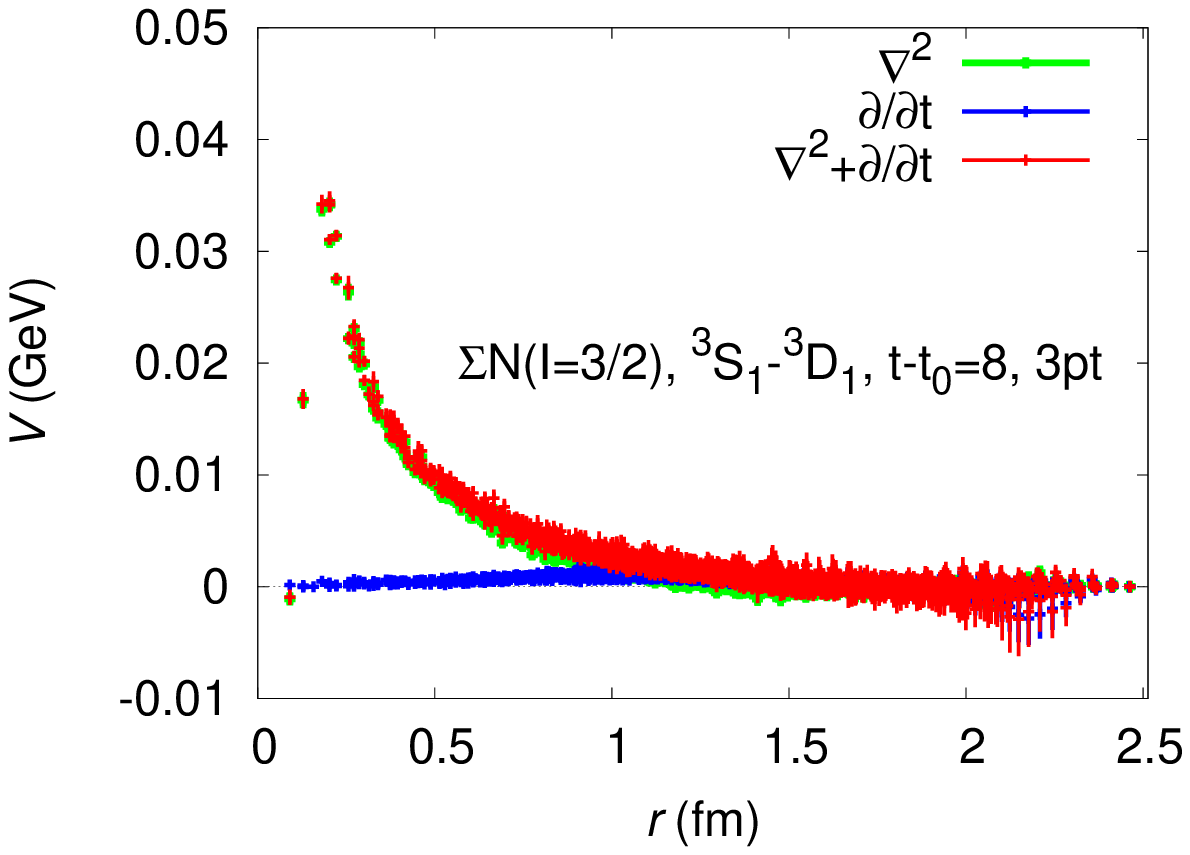}
 \end{minipage}
 \caption{
 Left:
 The central potential in the $^3S_1-^3D_1$ channel of the $\Sigma N$ system
 as a function of $r$. 
 Right:
 The tensor potential in the $^3S_1-^3D_1$ channel of the $\Sigma N$ system
 as a function of $r$.
 }
 \label{SN2I3_VCT_3E1}
\end{figure}

Figure~\ref{SN2I3_VCT_3E1} shows the central potential (left panel) 
and the tensor potential (right panel) of the $\Sigma N (I=3/2)$ system
in the $^3S_1-^3D_1$ channel. 
Due to the isospin symmetry,
this channel belongs solely to the flavor $\mathbf{10}$ representation without 
mixing with $\overline{\mathbf{10}}$ or $\mathbf{8}_a$

As seen form an enlargement in the inset of the left panel,
there is no clear attractive well in the central potential (red).
 An attractive well seen in the laplacian part  (green) at medium distances is
 cancelled by  a repulsive contribution in
the time-derivative part (blue) at the whole range.
This repulsive nature of the $\Sigma N (I=3/2, ^3S_1-^3D_1)$ 
central potential is consistent with the prediction from the 
naive quark model.\cite{Oka:2000wj} 
The tensor force is a little stronger that that of the $\Lambda N$ system but is still 
 weaker in magnitude than that of the $NN$ system.  
The time-derivative part gives  a negligible contribution. 

\section{Summary}

We have calculated the central and the tensor parts of the $\Lambda N$ and 
$\Sigma N(I=3/2)$ potentials using the improved method.  
Two  potentials in the $^1S_0$ channel, one is $\Lambda N$ and 
another is $\Sigma N(I=3/2)$,  are very similar to each other, 
since both sectors have a common $\mathbf{27}$ representation in the flavor SU(3) 
and the SU(3) breaking effect is still small in the present calculation. 
Both potentials have an attractive well, so that they 
give attractive interactions at low energy.

In the $^3S_1-^3D_1$ channel,
while the central $\Lambda N$ potential shows attraction at low energy with an attractive well,
the central $\Sigma N(I=3/2)$ potential  is repulsive at all distances.
Tensor potentials for both $\Lambda N$ and $\Sigma N(I=3/2)$ systems in this work
are rather weak. 

In this report, 
we have extracted single-channel potentials only without 
coupled-channel analysis between 
$\Lambda N$ and $\Sigma N$.
though the coupled-channel potentials can be also extracted in principle\cite{Aoki:2011gt,kenjisLAT2011}.
The author had postponed such  a coupled-channel analysis to future publications,
since he had to spend his time to recover in various parts of his life and research environment 
from the massive earthquake in East of Japan on March 11th, 2011. 
Further analyses including coupled-channel potentials are now
in progress. 
%


\acknowledgments

{%
The author would 
like to thank  
CP-PACS/JLQCD collaborations, 
PACS-CS Collaboration and 
ILDG/JLDG\cite{ILDGJLDG} 
for 
allowing us to access the full QCD gauge configurations, 
and 
Dr.~T.~Izubuchi 
for providing a sample FFT code.
}
The author also thank maintainers of \verb|CPS++|\cite{CPSPLUSPLUS}. 
Calculations in this report have been performed 
by using the Blue Gene/L computer 
under the ``Large scale simulation program'' 
at KEK (No. 10-24). 
He would also like to thank Dr.~S.~Ejiri, Dr.~K.~Itahashi 
and 
Advanced Meson Science Laboratory of RIKEN Nishina Center 
for providing a special computer resource, and 
Prof.~H.~Tamura and Dr.~T.~Koike for generous support 
in recovering the research environment after 
the massive earthquake in East of Japan on Math 11th. 
The author is supported by the Global COE Program for 
Young Researchers at Tohoku University. 
This research was supported in part
by Strategic Program for Innovative Research (SPIRE),
the MEXT Grant-in-Aid,  
Scientific Research on Innovative Areas 
(Nos. 21105515, 20105003).


\end{document}